\documentclass[reprint,
 amsmath,amssymb,aps,
]{revtex4-1}
\usepackage{graphicx,subfigure,color}
\usepackage{dcolumn}
\usepackage{bm}
\usepackage{hyperref}
\usepackage[mathlines]{lineno}
\newcommand{\be}{\begin{equation}}
\newcommand{\ee}{\end{equation}}
\newcommand{\bea}{\begin{eqnarray}}
\newcommand{\eea}{\end{eqnarray}}
\newcommand{\beq}{\begin{equation}}
\newcommand{\eeq}{\end{equation}}
\newcommand{\beqa}{\begin{eqnarray}}
\newcommand{\eeqa}{\end{eqnarray}}

\newcommand{\vev}[1]{\langle #1 \rangle}

\newcommand{\Tr}{{\rm Tr}\,}

\begin{document}

\title{Holographic Abelian Higgs model and the Linear confinement}
\author{Eunseok Oh and Sang-Jin Sin}
\email{lspk.lpg@gmail.com, sangjin.sin@gmail.com}
\affiliation{Department of Physics, Hanyang University, Seoul 04763, Korea.}  \date{\today}%
\begin{abstract} 
We consider the holographic abelian  Higgs model and show that,  in the absence of the scale symmetry breaking effect, chiral symmetry breaking  gives   linear confinement where the slope is given by the value of the chiral condensation.   The model can be considered either as the theory of superconductivity  or as the axial sector  of QCD depending on the  interpretation of the charge.  
We  also  provided a few models with linear confinement.  
\end{abstract}
\keywords{Abelian Higgs model, Holography, chiral symmetry}
\maketitle

\section {Introduction} 
One of most spectacular phenomena provided by the strong interaction is  the stringy structure  out of  systems  of particles, and the leading guidance in strong interaction has been the chiral symmetry. 
The string theory   was born by the observation  \cite{Veneziano:1968yb,Mandelstam:1974pi} that a linear relation between the mass squared and  spin in the data of  hadrons, the Regge trajectory, can be realized by the  spectrum of a vibrating string. 
Naturally,   in the era of the AdS/CFT \cite{Maldacena:1997re,Witten:1998qj,Gubser:1998bc}, two of the  leading   questions in its application were how to utilize the chiral symmetry\cite{erlich2005qcd,Karch:2006pv} in the new context  and whether  holography  can produce the linear trajectory for quantum chromodynamics (QCD). 

 When the vacuum expectation value of $\bar q q$ is non-zero, chiral symmetry is broken and  QCD is in the confining phase. Therefore  these two  are likely  related  and therefore linear Regge trajectory and  the chiral symmetry breaking should be so also. 
However  this has not been so clear even in the AdS/QCD era not alone in the perturbative field theory period, although  the linear spectrum for the vector meson sector was given in  \cite{Karch:2006pv} using a dilaton configuration.

In this paper,  we will point out that the holographic abelian Higgs model has linear confining spectrum. That is,   the chiral symmetry breaking    is enough to establish the Regge trajectory in the axial sector of the QCD. We consider  one flavor case for simplicity so that the theory become an abelian gauge theory.  {Notice that even for the multi-flavor case, the non-linearity due to the non-abelian structure is  irrelevant in leading order   discussion  on the spectrum and transports, which ensures that the linearity of the spectrum remains for multi-flavor case.  }

The left and right handed quarks,  $q_L,q_R$,   have axial charge $-1$ and $+1$ respectively under the axial $U(1)$  global symmetry. 
Invoking the holographic principle  we have axial  gauge field   $A_\mu$   and  a complex scalar field  $\Phi$ which is  the  dual to the    $\bar q_L q_R$ of charge 2.  
Because the presence of pion indicates that the chiral symmetry is broken spontaneously, so is the promoted chiral $U(1)$  gauge symmetry.
Thus we are naturally lead to the abelian Higgs model where $\Phi$ plays the role of Higgs field. 
We will show below that this model has the linearly confining spectrum.

On the other hand, we will see that when there is a dilatonic effect like in the soft-wall model \cite{Karch:2006pv}  spontaneous chiral symmetry breaking(ChSB) will not give any significant contribution  to the Regge  slope, instead it can contribute to the Regge intercept. 
Namely the Regge slope is predominantly determined by the gluon condensation only and this explains why all the Regge slopes are the same.  
  This can be traced back to the fact that  the complex scalar field is dominated by the   dilaton configuration in the infrared regime. Therefore in this paper we do not use overall dilaton factor $e^{-\varphi}$, and will  give a phenomenological QCD model where spontaneous chiral symmetry is not disturbed  by dilaton and linear confinement is respected in all sectors.

We will also give a various models with linear confinement properties which may and may not be related to the real QCD, because our observation on  the stringy spectrum is very universal not necessarily  attached to the phenomena of QCD. 

\section{Strings   in  holographic  Abelian Higgs Model}
We start from the  canonical action  of the   gauge field $A_\mu$ 
and  the complex boson $\Phi$ in a fixed metric background. 
\begin{eqnarray} 
S= \int d^{d+1}x\sqrt{-g}\Big( -\frac14 F_{\mu\nu}^2  -|D_\mu\Phi |^2 -m_\Phi^2 |\Phi|^2 \Big), \label{action}
\end{eqnarray}
 where  ${D}_\mu= \nabla_{\mu}  -i qA_\mu $ is the covariant derivative in the $AdS_{d+1}$ of radius $L=1$ whose 
metric is 
\be
ds^{2}=(dz^{2}+\eta^{\mu\nu}dx^{u}dx^{\nu} )/z^{2},  \hbox{ with } \eta^{00}=-1.
\ee
Bulk mass  $m_{\Phi}^2$ is given in terms of the conformal dimension of the dual operator: $m_{\Phi}^2 =\Delta(\Delta-d)$. We will fix it such that $\Delta=2$, which  is   natural in  $d=2+1$ dimension. 
For  $d=3+1$,   we need to choose $m_\Phi^2=-4$ for 
$\Delta_{\bar q q}=2$, which  is  realized   at the left boundary value of conformal window of  $N_{f}/N_{c}$ \cite{Kaplan:2009kr}. Since we are applying the AdS/CFT in the confined phase at the low energy where the conformality is lost, the     boundary value 2 is better than  the  free fermion value 3. 
The    field equation then  gives  
\bea
\Phi &=& M_0 z + {M}z^2,  \hbox{ in AdS$_{4}$, } \label{Phi}\\
\Phi &=& M_0 z^{2}\ln z^{-1}+ {M}z^2,  \hbox{ in AdS$_{5}$. } \label{Phi}
\eea
which are   exact solutions of the scalar field equations in the probe limit.  
Since we   look  for dynamically generated gap, we set the source   
$M_{0}=0$ so that  $\Phi={M}z^2$. 

Now the Maxwell equation  is given by 
\be
\nabla^{\mu}F_{\mu\nu}=J_{\nu}
\ee
and for the real solution of $\Phi$, the current is simplified to 
the London equation similarly to the superconductivity,
\be
J_{\mu}=q^{2} \Phi ^{2}A_{\mu}. 
\ee
For the transverse components with $\vec{k}\cdot \vec{A}=0$, it can be rewritten as Schr\"odinger  equation \cite{Karch:2006pv} via  $\Psi=\frac{A_{x}}{z^{(d-3)/2}}$ : 
\bea
-\Psi''_{n} &+&V\Psi_{n}=E_{n}\Psi_{n}, \label{vectorE} \\
V &=& \frac{p^{2}-\frac14}{z^{2}} +q^{2}M^{2}z^{2}  \label{potential}  \\
 E_{n}&=&qM(4n+2p+2),  \label{eigenV}  
\eea
 with $ p= (d-2)/{2}$,  and  $E_{n}=\omega^{2}-k^{2}\equiv m_{n}^{2}$.
The corresponding wave functions are given by
	\begin{align}
    \Psi_n(z)=N e^{- \frac{1}{2}qMz^2} z^{p+\frac12} L_{n}^{p}(qM z^2), 
	\end{align}
	where $N=\sqrt{\frac{2 n! (qM)^{p+1}}{(n+p)!}}$.
The mass spectrum is 
\be 
m_{n}^{2}=4qM(n+d/4). \label{stringSpec}
\ee 
For tensor with rank $s$, there are a few possible models according to the permutation symmetry of the index and gauge symmetry of the theory.  
As we have shown in the appendix,  some of them has  spectrum 
\bea 
m_{n,s}^{2}&=& 4qM(n+s-1+\frac d 4). \label{Sspec} 
\eea 
 For $d=4$, these results coincide with those of 
 \cite{Karch:2006pv},  where  vector meson spectrum  was discussed using  the  dilaton.  Notice that here  we did not use the dilaton.
The reason for such coincidence is just  because the equations of motion of two models turns out to be the same when they are expressed in Schr\"odinger form in spite of the difference in the degree of freedom.
However, this is because 
we use the scalar solution in the probe limit at zero temperature. When we consider the effect of  the finite temperature   or back reaction or chemical potential, the difference will be manifest.  
 For $d=3$, the $1/z^{2}$-potential is accidentally cancelled, but the spectrum is still given by above formula because we need to impose the boundary condition $A_{\mu}=0$ at the boundary of AdS. 
 
 For general spin $s$, we need to choose the mass term of the higher spin fields properly to get 
eq.(\ref{Sspec}).  
 That   spin dependent mass is necessary for the   spectral formula has been  known from the original paper \cite{Karch:2006pv}  but has not been so clear.  Notice  that  in string theory the action encodes all the spin simultaneously while 
in field theory the action for each spin should be considered one by one.   
  Now how  to add up such spin   dependent field theories to  describe the holographic image of the bulk fundamental string? While the kinetic terms are canonical,  the mass term and interaction term of spin $s$ excitation are  ambiguous. 
We suggest that reproducing the linear spectrum can be used as a guiding principle to determine them especially if our purpose is to describe a theory whose spectrum follows Regge trajectory. Then statement is that,  for any spin of given symmetry, there exist a choice of mass term such that the resulting spectrum is given by eq.(12). 
 

Notice that 
the spectrum is linear in both spin $s$ and   vibrational quantum   number $n$ and therefore the model has a spectrum of open string whose string tension is 
  \be
  T=1/(2\pi \alpha'), \quad \hbox{with }    \alpha'= 1/{(4qM)}.
\ee 
 In $M\to 0$ limit, that is, in the tensionless limit,    the whole tower of string spectrum is reduced to that of a massless particle. 
 
 Although we considered the only abelian theory, the same spectrum will be obtained for nonabelian theories. This is because holographic spectrum analysis  depends only on the quadratic part of the field variation's action.  Therefore,   when we consider SU(N) and we perturb around 0 background gauge field, the  non-linear terms induced by the non-abelian-ness does not affect the spectrum. Therefore Non-linear chiral dynamics in holography will also give linear spectrum.  That is,  our spectrum is the same as  that of the 2+1 dimensional version of   EKSS model \cite{erlich2005qcd} if boundary condition is the same. 
In fact, the important  difference is soft wall boundary condition (BC) installed by      
 scalar  condensation. The authors of  \cite{erlich2005qcd}  did not get linear spectrum  because they assumed the hard wall BC.   
 In \cite{Karch:2006pv}, the authors introduced softwall by hand using the dilaton dressing which is not supported by an equation of motion. 
Then in terms of the QCD,  what we did is to use the action of the hardwall model but install the soft wall BC by scalar condensation.

 \section{ Physical meaning of stringy excitations} 
 So far, we  have seen  that   the abelian Higgs model considered as the axial part of the QCD  has a linear spectrum. 
 In our field theory description, we have seen that 
each spin $s$ excitation of the   string in AdS, which  we called  'spin s particle in AdS', creates a tower of linear spectrum in the boundary and   summing all of them with different $s$ should be   interpreted as the string spectrum. 
    

What is the origin of the string?  AdS/CFT is the equivalence  of the gauge theory with string theory in AdS. 
It has also been  well accepted that the fundamental string in the bulk of the AdS geometry   is   dual to the color electric flux. 
Such duality is holographic statement of the Mandelstam- 't Hooft (MT) duality \cite{Mandelstam:1974pi,tHooft:1975yol,Peskin:1977kp} which conjectured that QCD vacuum  
realizes a dual superconductivity:  just as magnetic flux is confined in superconducting vacuum electric flux is confined in the vacuum.
 It would be good if we can understand this more explicitly.
  
Notice that  the abelian Higgs  model has long been identified as  the holographic theory of  superconductivity\cite{Gubser:2008px,Hartnoll:2008vx}  in 2+1 dimensional system, although     
 it also has   an interpretation as a theory of superfluidity \cite{Adams:2012pj}.  
 
 Notice that both the color flux tube and the stringy spectrum we obtained  are in the boundary theory. Therefore it is natural to identify our   spectrum obtained from the abelian Higgs model  with  the string  of confined color flux.  As we described earlier, the reason why U(1)  theory captured the confinement dynamics is because   only quadratic fluctuation is relevant in the spectral analysis: the non-abelian nature 
of SU(N) theory is washed out together with the non-abelian-ness created by the former.  
Notice also that the abelian Higgs model in AdS  was obtained by promoting the global chiral symmetry in the boundary.
Therefore, the chiral symmetry   describes 
the   confinement dynamics through holography when chiral condensation $M$ is assumed.  
 
The fact that both the theory of confining dynamics and theory of  superconductivity 
are described by the same  theory, the holographic abelian higgs model, can be considered  as an explicit   demonstration  of the Mandelstam- 't Hooft duality, which is nothing more than the statement that the way QCD vacuum works is mimicking the superconductivity. Here we are saying that they are not just mimicking each other: they are described by the same theory.

 \section{Two competing orders in QCD: scale and chiral symmetry breakings}
 So far we have not considered scale symmetry breaking. We now consider how including it can change the behavior of the theory. In fact, one of the important mechanism of mass generation in the QCD is the scale symmetry breaking.  The dilaton field has been usually considered as the dual of the gluon operator. 
In ref. \cite{Karch:2006pv} of the softwall model,  the  dilaton factor  $e^{-\varphi}$.  In this paper, we identify $\varphi$ as the square root of the gluon operator:  
$
 \varphi^{2} \sim \Tr{F_{\mu\nu}F^{\mu\nu}}
 $.  { Since we do not want to modify the AdS metric in Einstein frame, we   determine the dilaton in the AdS background.}
 Then $\varphi$ is of dimension 2 and we can write down the  bulk action of it using the bulk mass $m^{2}_{\varphi}=-4$. Setting the source part of $\varphi$ to be zero  as before, we get 
 \be 
 \varphi=G z^{2}, \quad \hbox{with  } G^{2}= {\vev{  \Tr{F_{\mu\nu}F^{\mu\nu}}}}
 \ee
 Now  the action of the softwall model is 
\be 
S= -\frac14 \int \sqrt{g} e^{-\varphi}(F^{2}_{A} +F^{2}_{V} +4| D_{A}\Phi|^{2}).
\ee
The Schr\"odinger form of the equation of transverse component of $V_{i}, A_{i}$ is $-\psi'' +V\psi =m_{n}^{2}\psi$ with  
\bea
V&=& G^{2}z^{2}+\frac{{3/4} }{z^{2}}
\hbox{ for vector } V_{i},  \\
&=& G^{2}z^{2}+\frac{1+|\Phi|^{2}-{1/4}}{z^{2}},  \hbox{ for  axial vector } A_{i} 
\eea
where 
$\psi={\cal A}_{\perp}/(\sqrt{z}e^{Gz^{2}/2})$  with ${\cal A}=V_{i},A_{i}$.
If the behavior of $\Phi = Mz^{2}$     were maintained, 
the Regge slope would be $ 4\sqrt{G^{2}+q^{2}M^{2}}$.  
However,  it can not be so:  the equation for the chiral scalar $\Phi$ in the presence of $\varphi$ is
\be
-\Phi''(z) +\big(\varphi'(z)+\frac3z \big) \Phi'-4\frac{\Phi}{z^{2}}=0.
\ee
Notice that the behavior of  $\Phi$ in large $z$ 
is dominated by $\varphi=Gz^{2}$, because   $\varphi'>>3/z$ there. Therefore asymptotic behavior of $\Phi$ is either 
   $\Phi \simeq \exp({Gz^{2}})$ or  
   $\Phi \simeq M_{1}\exp({-1/Gz^{2}})$ with dimensionless parameter $M_{1}$.   
 Since we should take the finite solution, we have 
 \be 
 \Phi 
 \simeq M_{1} \hbox{ for } z \to \infty.
 \ee
For large quantum number $n$, 
\be
m_{n}^{2}=  G(4n+2\sqrt{1+M_{1}^{2}}+2).
\ee   
For small $n$, configuration of $\Phi$ can make a small   non-linear component to the trajectory. 

Notice that  the  chiral symmetry breaking, although 
its breaking is  spontaneous, does not contribute to the Regge slope, so that the Regge slope is determined only by the scale symmetry breaking scale. 
 Indeed, the Regge slopes of all the meson family are the same and the we point out that this makes the softwall model  explain  why this is so. 
The chiral symmetry breaking contributes to Regge intercept by the parameter $M_{1}$, which is expected to be zero    in the limit of chiral symmetry restoration. This explains why the vector and axial vector mesons will be the same, which is another phenomenological fact. 

In summary, both chiral symmetry breaking and  non-trivial dilaton configuration discussed in this section  are natural ways to introduce a physical scale.  
The issue here was whether  two mechanism can co-exist  or compete. For the former case,    we would  have two independent scales in QCD.  
What we found here is that interestingly they compete and only one mechanism  survives and as a consequence we have only one scale .  
 
\section{Other models with linear confinement } 
In the rest of this paper,   we provide other  models {\it without} the over all dilaton factor $e^{-\varphi}$ yet  having linear Regge trajectories   for the future   model building for QCD and condensed matter.  
In the confined phase, we should   treat   particle of each  spin individually.
 \bea
 S &=&  \sum_{s\geqslant 2} S_{V,s}+S_{A,s},  
 \eea
where index ${s}$  is  for   spin $s$. 
  
The vector meson can not couple to $\Phi$ because  $\Phi$ does not have the vector charge.  
Usually the dilaton, the Goldstone boson of the scale symmetry, is introduced as  a real massless scalar which is dual to the gluon   operator  
$\Tr G^{2}_{\mu\nu}$ whose non-zero vacuum expectation value breaks the scale symmetry.  For our purpose we identity it as a square root of the gluon operator. 
It should couple to the vector meson otherwise the latter will be massless.  
The action of the vector meson is $S_{V} =  \int d^{d+1}x \sqrt{-g} {\cal L}_{V}$ with
\begin{eqnarray} 
 {\cal L}_{V} =   -\frac14 F_{V}^2 -\frac12 \nabla_\mu\varphi  \nabla^\mu\varphi  
- g^{2}\varphi^{2} V_{\mu}V^{\mu}  
 \label{actionV}
\end{eqnarray}
The (square root of) dilaton has following solution $\varphi= Gz^{2}$ as before.
The  equation for the transverse  vector meson is still given by the Schr\"odinger Eq. (\ref{vectorE}) with $qM$ replaced by $gG$.
 Therefore the  spectrum of the vector meson is again a linear tower   given by 
 \be
 m_{n,vector}^{2}={4gG}(n+d/4). \label{stringSpecV}
 \ee 
 If we have added $-m_{V}^{2} V_{\mu_{1}\cdots\mu_{s}}^{2} $ term to the Lagrangian of spin $s$ vector meson, the spectrum would change to 
\bea 
m_{s,n}^{2}&=& 4gG\big(n + \frac{p_{V}+1}{2} \big). \label{Sspec3} 
 \eea
 with $p_{V}= 2\sqrt{ \big(s-1+\frac{d-2}{4}\big)^{2}+m_{V}^{2}}$.
 To fit the data for $\rho$ meson with $d=4$, $s=1$, we can take $p\simeq -1$ which can be done 
 most naturally  by setting $m_{V}^{2}=0$. That is, for the   phenomenology, it is better not to introduce the bulk mass of the vector meson.

%

\vskip.2cm

\noindent {\it Gluon condensation and axial mesons:}
%
The anomaly of the $U(1)_{A}$ can be considered as a part of the spontaneous breaking of the axial symmetry and we should open the possibility that the promoted bulk gauge invariance can be broken explicitly at the bulk level, because the bulk theory should include the quantum dynamics of the boundary theory at the classical level.   
This implies that Axial symmetry could   have been further broken by  
adding the bulk mass term $-m_{A}^{2} A_{\mu_{1}\cdots\mu_{s}}^{2} $   to the Lagrangian of spin $s$ axial vector meson. 
 
Then the spectrum would change to 
 \bea 
m_{s,n}^{2}&=& 4qM\big(n + \frac{p_{A}+1}{2} \big). \label{Sspec4} 
 \eea
 with $p_{A}= 2\sqrt{ \big(s-1+\frac{d-2}{4}\big)^{2}+m_{A}^{2}}$. 
 Again when one consider the equation of motion in the Schradinger form, they become silmilar to the model having the dilaton softwall model, and according to refs. \cite{Anisovich:2000kxa, Huang:2007fv} $4qM=1.25(GeV)^{2}$ and $m_{A}=0.5$ can fit the data well.  

It could   have been broken even further  by   dilaton coupling   $- \varphi A_{\mu}A^{\mu}$.
However,  then the spectrum is changed by $qM\to\sqrt{ (qM)^{2} +(gG)^{2}}$ so that 
the slope of Regge trajectory of the axial vector is   bigger than that of vector meson, which is not consistent  with the data  in \cite{Anisovich:2000kxa}.  Therefore we do not add dilaton coupling of the axial meson. 
To make two slope equal, we need 
\be
gG=qM.
\ee
However, this is not a good consequence for the QCD, because it  means
 a  fine tuning is necessary for the universality of the Regge slope. 
\vskip .2cm

 \noindent   {\it Glue ball spectrum:} 
 To understand the color confinement, it is good idea to look at the  
    behavior of a gauge invariant version of color fields, say ${\cal O}=\Tr (G_{\mu\nu})^{n} $,  under the gluon condensation. 
Let $\phi$ be the scalar field in the bulk which is dual to the scalar field $\cal O$ 
of dimension $\Delta$. Then the dynamics of $\phi$ 'inside' the bag  can be studied by   
	\begin{align}
	S_\phi =\frac12\int d^{5}x\sqrt{-g}(-\nabla^\mu\phi\nabla_\mu\phi
	-m^2\phi^{2}-g_{S}^{2}\varphi\phi^{2} )
	\end{align} 
	Using the   solution  $\varphi=G^{2}z^{4}$ as before,  the {Schr\"odinger form} of the scalar equation is given by Eq.(\ref{vectorE}) with $p_{S}^{2}=m^{2}+4$,
and the scalar meson spectrum is given by 
 \be
 m_{n,scalar}^{2}={4g_{S}G}\big(n+\frac12(p_{S} +1) \big).
\label{stringSpecV}
 \ee  
 The linear spectrum of the glueball is interesting but 
 what is more important for us here is the behavior of the wave function 
	Eq.(\ref{potential}) which says that the color flux outside the bag, $z>z_{m}$ is 
	exponentially suppressed,
 proving the color confinement within the bag under the presence of the gluon condensation. 
 
 Notice that in many of our models, we need to choose the bulk  mass of the theory properly to get the promised combination $n+s$. 
That   spin dependent mass is necessary for the   spectral formular  has been  known from the original paper \cite{Karch:2006pv}  but has not been clear so far.  Notice  that string theory encode all the spin simultaneously while 
in field theory the action for each spin should be considered one by one.   
  Now how we add up such spin   dependence field theories to  describe the holographic image of the bulk fundamental string? While the kinetic terms are canonical,  it is not surprising to have ambiguities in the mass term of spin $s$ excitation. 
We suggest that reproducing the linear spectrum can be used as a guiding principle to determine them.

	\vskip.2cm

 \section{Discussion}
    
 We finish the paper by summary and a few remarks.
   First one may ask the problem of blowing up of the scalar solution in  IR region ($(z\to \infty)$). This is precisely  the problem of probe approach where we assume that the gravity background is fixed as AdS.  There is a known resolution to this: in reality, the back-reaction of AdS will either create horizon, a natural IR cut off,  or smooth out the solutions.  
  Whether the probe solution is useful or not depends on what we do with it. If we evaluate the thermodynamic quantity, we would fail. But for the spectrum, the background  will be useful because true solution will be similar to the probe solution away from the singular region ($z\to \infty$), which is  forbidden for the wave function of excitations anyway:  we are looking for equation of motion of the vector field's perturbation, which was shown to be written (after a change of variable ) as a Schr\"odinger equation whose potential contains   $z^{2}$ term with $z=1/r$. 
Such configuration  provide a softwall providing the barrier   so that the wave function will not penetrate to the IR region $z\to \infty$.   
This is an effective way to cut out the IR regime and justify the use of the probe solution for the problem  of spectrum.   One can show that   when we consider the back reaction, a horizon is developed and as the horizon grows  the the potential's $z^{2}$ regime will retract. The potential  will not grow like $z^2$ indefinitely but collapse to to $-\infty$ at the horizon, so that  the higher quantum number of the linear spectrum will be deformed and disappear. Details of such effect is a complicated correction to the simple  phenomena describe here.

Next, 
in a theory where   symmetry breaking does not enter, chiral symmetry breaking can contribute to the slope of axial vector meson.
But when scale symmetry breaking comes with coupling of overall 
$e^{-\varphi}$ coupling, like  soft-wall model, the theory changes its face: the chiral symmetry breaking effect  is eaten  by that of the scale symmetry breaking  and  does not contribute to the Regge slope. 
However, the latter can contribute to the intercept of Regge trajectory and the mass of the axial vector meson.

Two models with and without the overall dilaton $e^{-\varphi}$ coupling, have pros and cons: 
 The model without such coupling makes the Mandelstam-'t Hooft duality is manifest but  we should  break vector gauge symmetry   explicitly  to give  vector meson mass  and universlaity of the Regge slope  need fine-tuning. 
On the other hand, the original soft-wall model does not have a manifest duality but treat the vector and axial vector sector symmetrically so that the universlaity of the Regge slope could  be shown. 

 We identified the origin of the Regge slopes as the condensation of order parameter that controls the  symmetry breaking scales.  
 The linearity of the Regge trajectory  is generated because the potential is the same form as that of 3-dimensional isotropic harmonic oscillator where $z$ play the role of the radial coordinate.  The 'centrifugal' term  is due to the confining gravity of the AdS space, while  quadratic potential is by gluon and chiral condensations.
The latter provides infinite ``soft wall''  and it can be attributed as a nature of the vacuum of such condensation. 
Our results suggest that the color confinement and the Regge slope is consequence of gluon condensation.  Therefore by measuring the Regge slopes,  we can determine it,  but  we can not determine the chiral condensation so easily.

\appendix
\section{Models with higher rank tensors}
In the first two subsections of this appendix,  we study models with diffeomorphism invariance but without gauge invariance. In the final subsection we study 
 the theory with  gauge invariance as well as diffeomorphism invariance. 
\subsection{Rank-s totally Antisymmetric Tensor \\without gauge symmetry}
 We may start with field equation
	\begin{align}
	\frac{1}{\sqrt{-g}}\partial_{\mu}(\sqrt{-g}g^{\mu\nu}g^{\alpha_{1}\beta_{1}}...g^{\alpha_{s}\beta_{s}}\partial_{\nu}A_{\beta_{1}...\beta_{s}})\\ \nonumber
	=(\Phi^{2}+{m^{2}_{A}})g^{\alpha_{1}\beta_{1}}...g^{\alpha_{s}\beta_{s}}A_{\beta_{1}...\beta_{s}}
	\end{align}
	With axial gauge choice $A_{zx_{1}...x_{s}}=0$, the equation of motion of $A_{x_{1}...x_{s}}:= B e^{ i(k\cdot x-\omega t)} $  takes the form,   
	\begin{align}
	-z^{-\alpha}\partial_{z}(z^{\alpha}\partial_{z}B)
	+(\Phi^{2}+m^{2}_{A})z^{-2}B= m_{n}^{2}B, 
	\end{align}
	where $\alpha=-d+2s+1$ and $m_{n}^{2}=\omega^{2}-k^{2}$.
Using the	identity
	\begin{align}
	\partial_{z}(z^\alpha\partial_{z}B)&=z^{\alpha/2}\Big(\phi''-\frac{(\frac{\alpha-1}{2})^{2} -\frac14}{z^2}\phi \Big)
	\end{align}
	with $B=z^{-\alpha/2} \phi$, 
	we get 
	\begin{align}
-\phi''+(\frac{p^2+m^2_{A}-\frac{1}{4}}{z^2}+M^2z^2)\phi =m_{n}^{2}\phi
	\\
E_{n,s}=M(4n+2\sqrt{p^2+m^2_{A}}+2),
	\end{align}
with 	$p^{2}=(s-\frac{d}{2})^{2}$.   Then  the desired spectrum 
\be
E_{n,s}=M(4n+4s-4+d), 
\ee
can be obtained for $m^2_{A}=3(s-1)(s+d-3)$.

 \subsection{Rank-s totally Symmetric Tensor \\without gauge symmetry}
For the same gauge choice and the variable, the 
	equation of Motion is 
	\begin{align}
V(z)=\frac{(\frac{d}{2})^2+s-\frac{1}{4}+m^2_A}{z^2}+M^2 z^2
	\end{align}
The spectrum 
\be
E_{n,s}=M(4n+4s+d-4), 
\ee
can be obtained  if $m^2_{A}= (2s-3)(2s-3+d)-s$.

If there were overall dilaton  factor $e^{\varphi}$ with $\varphi=Mz^{2}$ in the action,
	\begin{align}
	V(z)=\frac{(\frac{d}{2})^2+s-\frac{1}{4}+m^2_A}{z^2}+M^2 z^2+d-2
	\end{align}
		then we can have 
\be
E_{n,s}=M(4n+4s-4+d), 
\ee
by choosing  $m^2_{A}=4s^2-9s+4 -d^{2}/4$.    

	\subsection{ Higher spin Theory}
	In \cite{Karch:2006pv}, the  rank-s totally symmetric Tensor with gauge symmetry $A_{\mu_{1} ...\mu_{s}} \to A_{\mu_{1} ...\mu_{s}}+\nabla_{(\mu_{1}}\xi_{\mu_{2} ...\mu_{s})}$ was identified as the spin s theory.   The residual gauge transformation which leaves $A_{z\mu_{2} ...\mu_{s}} $ invariant \cite{Karch:2006pv} is determined by 
	\be
	\nabla_{(\mu_{1}}\xi_{\mu_{2} ...\mu_{s})}=0.
	\ee
	Using $
	\Gamma^{\mu}_{z\mu}=-1/z, \Gamma^{z}_{ii}=1/z,  \Gamma^{z}_{tt}=-1/z,$
	 we get 
	  \be
	  \partial_{z}\xi +\frac{2s-2}{z} \xi=0,
	  \ee 
namely  $z^{2s-2}\xi_{x}(z,{\bf x} ):={\tilde \xi}_{x_{1},...,x_{s}}({\bf x})$ is $z$ independent. 
	Introducing the scaled variable 
	${\tilde A}_{x_{1},...,x_{s}}: =z^{2s-2}A_{x_{1},...,x_{s}} $, the residual gauge transformation in terms of the tilde variable is nothing but the shifting: ${\tilde A}_{x_{1},...,x_{s}}\to {\tilde A}_{x_{1},...,x_{s}}+{\tilde \xi}_{x_{1},...,x_{s}}$. 
	The action can be written as 
	\be
	S=\int z^{\alpha}e^{\varphi}[(\partial_{\mu}{\tilde A}_{x_{1},...,x_{s}})^{2} ],\label{A3}
	\ee
where with $\varphi=Mz^{2}$ and  $\alpha=-(1+d) +2(s+1) -2(2s-2)=4-d+1-2s$. 	In other words, the action should be designed such that Eq.(\ref{A3}) is hold using covariant derivatives.  
Now using the methods which is by now familiar, we have  		
\be
E_{n,s}=M(4n+4s-4+d).
\ee
as it was described in \cite{Karch:2006pv} for $d=4$.  

One should notice that the mass term is not invariant and therefore the invariance under the residual gauge transformation should determine the mass  \cite{Karch:2006pv} to be $m_{A}^{2}=s^{2}-s-4$. For the same reason, 
the naive scalar coupling  term such as $\Phi^{2} ({\tilde A}_{x_{1},...,x_{s}})^{2}$ is not allowed. 

 \begin{acknowledgements}
 We thank Youngman Kim,  Yongseok Oh,   Junchen Rong  and especially to Mannque Rho for useful discussions.  This  work is supported by Mid-career Researcher Program through the National Research Foundation of Korea grant No. NRF-2016R1A2B3007687. We  are also grateful for the support of APCTP during the 2018 focus workshop. 
\end{acknowledgements}

\bibliographystyle{JHEP}
\bibliography{Refs_scalar.bib}

\providecommand{\href}[2]{#2}\begingroup\raggedright\begin{thebibliography}{10}

\bibitem{Veneziano:1968yb}
G.~Veneziano, \emph{{Construction of a crossing - symmetric, Regge behaved
  amplitude for linearly rising trajectories}},
  \href{http://dx.doi.org/10.1007/BF02824451}{\emph{Nuovo Cim.} {\bf A57}
  (1968) 190--197}.

\bibitem{Mandelstam:1974pi}
S.~Mandelstam, \emph{{Vortices and Quark Confinement in Nonabelian Gauge
  Theories}}, \href{http://dx.doi.org/10.1016/0370-1573(76)90043-0}{\emph{Phys.
  Rept.} {\bf 23} (1976) 245--249}.

\bibitem{Maldacena:1997re}
J.~M. Maldacena, \emph{{The Large N limit of superconformal field theories and
  supergravity}},
  \href{http://dx.doi.org/10.1023/A:1026654312961}{\emph{Int.J.Theor.Phys.}
  {\bf 38} (1999) 1113--1133}, [\href{http://arxiv.org/abs/hep-th/9711200}{{\tt
  hep-th/9711200}}].

\bibitem{Witten:1998qj}
E.~Witten, \emph{{Anti-de Sitter space and holography}}, {\emph{Adv. Theor.
  Math. Phys.} {\bf 2} (1998) 253--291},
  [\href{http://arxiv.org/abs/hep-th/9802150}{{\tt hep-th/9802150}}].

\bibitem{Gubser:1998bc}
S.~S. Gubser, I.~R. Klebanov and A.~M. Polyakov, \emph{{Gauge theory
  correlators from noncritical string theory}},
  \href{http://dx.doi.org/10.1016/S0370-2693(98)00377-3}{\emph{Phys. Lett.}
  {\bf B428} (1998) 105--114}, [\href{http://arxiv.org/abs/hep-th/9802109}{{\tt
  hep-th/9802109}}].

\bibitem{erlich2005qcd}
J.~Erlich, E.~Katz, D.~T. Son and M.~A. Stephanov, \emph{Qcd and a holographic
  model of hadrons}, {\emph{Physical Review Letters} {\bf 95} (2005) 261602}.

\bibitem{Karch:2006pv}
A.~Karch, E.~Katz, D.~T. Son and M.~A. Stephanov, \emph{{Linear confinement and
  AdS/QCD}}, \href{http://dx.doi.org/10.1103/PhysRevD.74.015005}{\emph{Phys.
  Rev.} {\bf D74} (2006) 015005},
  [\href{http://arxiv.org/abs/hep-ph/0602229}{{\tt hep-ph/0602229}}].

\bibitem{Kaplan:2009kr}
D.~B. Kaplan, J.-W. Lee, D.~T. Son and M.~A. Stephanov, \emph{{Conformality
  Lost}}, \href{http://dx.doi.org/10.1103/PhysRevD.80.125005}{\emph{Phys. Rev.}
  {\bf D80} (2009) 125005}, [\href{http://arxiv.org/abs/0905.4752}{{\tt
  0905.4752}}].

\bibitem{tHooft:1975yol}
G.~'t~Hooft, \emph{{Gauge Theory for Strong Interactions}},  in \emph{{New
  Phenomena in Subnuclear Physics: Proceedings, International School of
  Subnuclear Physics, Erice, Sicily, Jul 11-Aug 1 1975. Part A}}, p.~0261,
  1975.

\bibitem{Peskin:1977kp}
M.~E. Peskin, \emph{{Mandelstam 't Hooft Duality in Abelian Lattice Models}},
  \href{http://dx.doi.org/10.1016/0003-4916(78)90252-X}{\emph{Annals Phys.}
  {\bf 113} (1978) 122}.

\bibitem{Gubser:2008px}
S.~S. Gubser, \emph{{Breaking an Abelian gauge symmetry near a black hole
  horizon}},
  \href{http://dx.doi.org/10.1103/PhysRevD.78.065034}{\emph{Phys.Rev.} {\bf
  D78} (2008) 065034}, [\href{http://arxiv.org/abs/0801.2977}{{\tt
  0801.2977}}].

\bibitem{Hartnoll:2008vx}
S.~A. Hartnoll, C.~P. Herzog and G.~T. Horowitz, \emph{{Building a Holographic
  Superconductor}},
  \href{http://dx.doi.org/10.1103/PhysRevLett.101.031601}{\emph{Phys.Rev.Lett.}
  {\bf 101} (2008) 031601}, [\href{http://arxiv.org/abs/0803.3295}{{\tt
  0803.3295}}].

\bibitem{Adams:2012pj}
A.~Adams, P.~M. Chesler and H.~Liu, \emph{{Holographic Vortex Liquids and
  Superfluid Turbulence}},
  \href{http://dx.doi.org/10.1126/science.1233529}{\emph{Science} {\bf 341}
  (2013) 368--372}, [\href{http://arxiv.org/abs/1212.0281}{{\tt 1212.0281}}].

\bibitem{Anisovich:2000kxa}
A.~V. Anisovich, V.~V. Anisovich and A.~V. Sarantsev, \emph{{Systematics of q
  anti-q states in the (n, M**2) and (J, M**2) planes}},
  \href{http://dx.doi.org/10.1103/PhysRevD.62.051502}{\emph{Phys. Rev.} {\bf
  D62} (2000) 051502}, [\href{http://arxiv.org/abs/hep-ph/0003113}{{\tt
  hep-ph/0003113}}].

\bibitem{Huang:2007fv}
S.~He, M.~Huang, Q.-S. Yan and Y.~Yang, \emph{{Confront Holographic QCD with
  Regge Trajectories}},
  \href{http://dx.doi.org/10.1140/epjc/s10052-010-1239-0}{\emph{Eur. Phys. J.}
  {\bf C66} (2010) 187--196}, [\href{http://arxiv.org/abs/0710.0988}{{\tt
  0710.0988}}].

\end{thebibliography}\endgroup

\end{document}